# INVESTIGATION OF THE STRUCTURE AND THERMODYNAMICS OF STAR-POLYMERS IN SEMI-DILUTE SOLUTION

F. Benzouine, M. Yassine, A. Derouiche, A. Bettachy, M. Benhamou

Laboratoire de Physique des Polymères et Phénomènes Critiques
Faculté des Sciences Ben M'sik, P.B. 7955, Sidi Othman, Casablanca, Morocco

**Abstract :** In this work, we consider a semi-dilute solution of identical star-polymers, made of attached flexible long polymer chains of the same polymerization degree N. We first compute the effective pair-potential between star-polymers. Such a potential results from the excluded volume forces between monomers. We show that this potential is logarithmic, below some known characteristic distance, $\sigma$, depending on the number of attached chains per star-polymer, f, and volume fraction of polymers, $\phi_0$. Beyond $\sigma$, the potential fails exponentially. Second, we investigate the structure and thermodynamics of these star-polymers. To this end, we use the *integral equation* method with the *hybridized-mean spherical approximation*. The numerical resolution of this equation gives the structure factor of the star-polymers, for various particle densities. Finally, the standard relationships between thermodynamics and structure enable us to deduce three physical quantities, namely the isotherm compressibility, pressure and internal energy, as functions of density.

## I. INTRODUCTION

Colloidal systems are often made of stabilized small particles, because they carry an electric charge. Their stabilization is then caused by the Coulomb force [1-3]. However, when the ionic concentration is increased, by addition of a salt or an electrolyte, the Coulomb force is then screened and replaced by a long-range Van der Waals attractive one. The latter generally originates from the fact that the particles possess a dipolar moment.

To prevent the particles flocculation, one way consists in introducing an adequately soluble polar head polymer chains or a copolymer with insoluble sequences. This constitutes the grafting phenomenon [4-9]. The result is that, two adjacent colloids clothed each by end-grafted flexible long polymer chains, repel each other due to the excluded volume forces between monomers.

Considerable simplifications occur when the particles are small enough, to be considered as *star-polymers*. Therefore, star-polymers provide a good model for studying statistical mechanics and thermodynamics of small colloidal particles, which are surrounded each by end-grafted polymer chains. From a topological point of view, star-polymers are typical branched macromolecules, possessing a central core from which many end-attached long linear polymer chains (or *branches*) emerge [10-12].

The stabilization of colloids with a grafted polymeric layer is the subject of many studies. The fundamental problem to solve is the knowledge of the nature of the effective force between particles. As we said above, such a force results from the excluded volume effect. The first theoretical work was due to Witten and Pincus (WP) [13]. The authors computed the repulsive force between two adjacent clothed particles, versus their center-to-center distance. They find that the force simply decays as the inverse of distance, with universal amplitude depending only on the number of branches per interacting star-polymer. To do calculations, they first assimilated these clothed particles to star-polymers [14], and used the popular model introduced by Daoud and Cotton (DC) [10], according to which, in dilute solution, each grafted chain or branch can be viewed as a sequence of growing spherical blobs.

The pioneered WP work stimulated much theoretical studies, dealt with good solvent of small molecules [12,15-21], $\theta$-solvent [22], high molecular-weight solvent [23], polydispersity [24], confinement [25] or chemical mismatch effect [26,27]. The used techniques ranged from scaling arguments to Renormalization-Group.

For the studies in relation with good solvents, particularly, the attention has been paid to star-polymers in dilute solution. This means that one has considered the problem of two star-polymers, only. To our knowledge, no theoretical work has been devoted to the effective force between star-polymers, in semi-dilute solution, where the excluded volume forces are screened. This is precisely the aim of the present work.

Our results are as follows. We first derive the expression of the pair-potential between interacting star-polymers. Contrary to dilute solutions, the latter explicitly depends on the monomer concentration. Second, with the help of the obtained effective potential, we investigate the structure and thermodynamics of the clothed particles by end-grafted polymer chains. Beside the pair-potential, we need a crucial object for determining the most physical properties, which is the *pair-correlation function*, $g(r)$. The latter is solution to the Ornstein-Zernike (OZ) integral equation [28]. But, this equation involves another unknown, that is the direct correlation function $c(r)$. Therefore, this necessitates a certain closure, that is, a supplementary relationship between these two correlation functions. Integral equation has been intensively used in modern liquid theory [28], and solved using some techniques, which are based on the analytical or numerical computation. In this context, various closures have been proposed, namely, the Percus-Yevick approximation [29], the hypernetted chain [30], the mean spherical approximation and its modification, that is the *hybridized-mean spherical approximation* (HMSA) [31]. Using this method, we compute the structure factor, which is the Fourier transform of the total pair-correlation function $h(r) = g(r) - 1$, for different values of particle density. The relations between the isotherm compressibility, the pressure and energy, and the pair-correlation function enable us to compute these three physical quantities, for the chosen particle densities.

The remaining of presentation proceeds as follows. In Sec. II, we compute the pair-potential. We investigate, in Sec. III, the structure and thermodynamics of the considered solution. We draw some concluding remarks in the last section.

## II. EFFECTIVE PAIR-POTENTIAL

Before the computation of the effective pair-potential, it will be instructive to recall some useful backgrounds concerning the conformation study of star-polymers in dilute and semi-dilute solutions.

### II.1. Star-polymers in semi-dilute solution

A star-polymer is characterized by two kinds of parameters N and f. The former is the polymerization degree of the attached chains, and the second, their number. Of course, the conformation of star-polymers depends on the value of the monomer concentration.

To fix ideas, we start by considering a single star-polymer of f branches, immersed in a good solvent. Thus, we are concerned with a very dilute solution. According to the DC model [10], an attached chain can be regarded as a sequence of growing spherical blobs. At a distance r from the centre of the star-polymer, the blobs cover a sphere of radius r. If $\xi(r)$ denotes the blob size at the considered distance, then, we can write

$$f.\xi^2 = 4\pi r^2 , \qquad (1)$$

Therefore, the blob size scales as [10]

$$\xi(r) \propto \frac{r}{\sqrt{f}} \ . \qquad (2)$$

This formula is valid only when one is above the core-radius [10] : $R_c \propto a\sqrt{f}$. Here, a means the monomer size. Below the length $R_c$, the blob size becomes of the order of a. This means that the f attached chains are *stretched* near the core.

On the other hand, at any distance from the centre $(r > R_c)$, the attached chains behave as in semi-dilute regime. Therefore, the local volume fraction of monomers, $\phi(r)$, is given by [32,33]

$$\phi(r) \propto \left[\frac{\xi(r)}{a}\right]^{-4/3} \ . \qquad (3)$$

Explicitly,

$$\phi(r) \propto f^{2/3} (r/a)^{-4/3} \ , \qquad R_c < r < R_G \ , \qquad (4a)$$

$$\phi(r) = 1 \ , \qquad r < R_c \ , \qquad (4b)$$

where $R_G$ represents the gyration radius of the isolated star-polymer, which can be measured in X-rays scattering experiment [32,33]. The corona-size $R_G$ can be obtained using the conservation law of the total mass

$$N.f = a^{-3} 4\pi \int_{R_0}^{R_G} dr\, r^2\, \phi(r) \ , \qquad (5)$$

together with relation (4a). Here, $R_0$ means the radius of the particle on which the f chains are attached. Straightforward calculations yield

$$R_G \propto a f^{1/5} N^{3/5} \ . \qquad (6)$$

We emphasize that the above formula makes sense only when the polymerization degree N exceeds the typical value $\sqrt{f}$, that is $N > \sqrt{f}$. This condition emerges from the fact that one must have $R_G > R_c$. Also, this same formula remains valid as long as the volume fraction of the solution $\phi_0$ is below some threshold, $\phi_0^*$, defined by

$$\phi_0^* = a^3 \frac{fN}{R_G^3} \propto f^{2/5} N^{-4/5} . \tag{7}$$

For $\phi_0 < \phi_0^*$ (dilute solution), the star-polymers behave as separated (swollen) *coils*, which avoid each other completely. At $\phi_0 = \phi_0^*$ (overlapping threshold), the coils begin to be densely packed. For $\phi_0^* < \phi_0 \ll 1$ (semi-dilute solution), however, the star-polymers overlap and the excluded volume is screened over distances much greater than the screening length $\xi_0 \propto a f^{1/2} \phi_0^{-3/4}$ [10]. Notice that, at the threshold, this length becomes of the order of the gyration radius $R_G$.

Now, assume that one is in semi-dilute regime, that is above the threshold $\phi_0^*$. The natural question to ask is about the impact of an increasing of the monomer concentration on the conformation of the star-polymers. Since the excluded volume is screened at high distances, and according to the DC image, an attached chain can be divided into two parts [10]. The outer part behaves as in semi-dilute solution, of volume fraction $\phi_0$, and can be considered as a sequence of *uniform blobs* of size $a \phi_0^{-3/4}$. The inner one, however, behaves as in dilute regime, for distances (from the centre) below some characteristic length, $\sigma$, manifestly smaller than $R_G$. This length can be obtained equating the size of the outer blob (at distance $\sigma$) $\xi(\sigma)$ to the blob size, that is $\sigma/\sqrt{f} = a\phi_0^{-3/4}$. This gives

$$\sigma \propto a f^{1/2} \phi_0^{-3/4} , \qquad \phi_0^* < \phi_0 \ll 1. \tag{8}$$

Thus, $\sigma$ depends on the number of branches per star-polymer f and the monomer concentration $c_0 = a^{-3} \phi_0$, and not on the polymerization degree N. Of course, the two lengths $\sigma$ and $R_G$ coincide at the threshold $\phi_0^*$. Combining relations (6), (7) and (8) yields the equivalent formula

$$\sigma = R_G \left(\phi_0 / \phi_0^*\right)^{-3/4} , \qquad \phi_0^* < \phi_0 \ll 1 . \tag{9}$$

For dilute regime, however the length $\sigma$ rigorously equals the gyration radius $R_G$, and we write

$$\sigma = R_G \ , \qquad \qquad \phi_0 < \phi_0^* \ . \qquad (10)$$

The schematic variation of the ratio $\sigma/R_G$ upon the dimensionless variable $\phi_0/\phi_0^*$ is depicted in Fig. 1. This curve traduces the natural decreasing of the length $\sigma$ with increasing monomer concentration.

Now, we have all ingredients to compute the effective pair-potential. This is precisely the aim of the following subsection.

**II.2. Effective pair-potential**

Assume that one is in semi-dilute solution and consider two star-polymers, which are at finite distance r apart. The existence of the excluded volume forces induces an effective pair-potential we want to compute.

The above discussions suggest that, for $r < \sigma$, the two star-polymers interact through a pair-potential as in dilute regime. For $r \gg \sigma$, the excluded volume is completely screened, and in principle, no mutual interactions are present. In the narrow region between the outer and inner parts of the two adjacent star-polymers, whose size is of the order of $2\sigma/\sqrt{f}$, we expect an exponential decay for the pair-potential.

With these considerations, the effective pair-potential takes the following form [18]

$$\frac{U(r)}{k_B T} = \frac{5}{18} f^{3/2} \times \begin{cases} -\ln(r/\sigma) + \left(1 + \sqrt{f}/2\right)^{-1} \ , & \text{for } r < \sigma, \\ \left(1 + \sqrt{f}/2\right)^{-1} (r/\sigma) \exp\left\{-\sqrt{f}\,(r-\sigma)/2\sigma\right\}, & \text{for } r > \sigma. \end{cases} \qquad (11)$$

Here, T is the temperature and $k_B$ the Boltzmann's constant.

Let us make some comments about the obtained pair-potential.

Firstly, we emphasize that the potential amplitude $(5/18)f^{-3/2}$ is not affected by an increasing of the monomer concentration. Indeed, such an amplitude is related to the small-distance behavior of the potential ; this corresponds to the inner region, where the solution is dilute. As for the constant term $\left(1+\sqrt{f}/2\right)^{-1}$ in the above expression, it is introduced to ensure the continuity and derivability properties of the potential at $r=\sigma$.

Secondly, contrary to dilute solutions, the interaction potential for semi-dilute ones depends on the monomer concentration, through the length $\sigma$, relation (8). But, it is independent on the molecular-weight of attached chains. This fact is inherent to semi-dilute solutions.

Thirdly, it is easy to see that, at fixed center-to-center distance, the pair-potential is shifted towards its lowest values, as the monomer concentration is increased. This is not surprising, since an increase of the monomer concentration implies a strong reduction of the excluded volume forces.

Finally, in Fig. 2, we report the reduced pair-potential $U(r)/k_BT$ versus the renormalized distance $r/R_G$, for 4 values of the reduced number density of particles $n/n^*$. For these curves, the number of branches f is fixed to the value $f=18$. Here, the characteristic value $n^*$ of the number density is $n^*=1/R_G^3$ [34]. The latter corresponds to the situation to have only one particle in the volume $R_G^3$. Notice the trivial relation $n=a^{-3}\phi_0/MNf$; thus, $n^*=a^{-3}\left(\phi_0^*/MNf\right)=1/R_G^3$. Here, M is the number of star-polymers in the solution, and $MNf$ the total number of monomers.

The effective pair-potential we obtained is the principal ingredient for the study of the structure and thermodynamics of a semi-dilute solution of star-polymers. This is the goal of the next section.

### III. STRUCTURE AND THERMODYNAMICS

To investigate the structure of the colloidal solution, made of clothed particles of high density $n>n^*$, we start from the OZ integral equation, satisfied by the total correlation function $h(\mathbf{r})=g(\mathbf{r})-1$ [28]. This integral equation involving the so-called direct correlation function $c(\mathbf{r})$, is given by

$$h(\mathbf{r})=c(\mathbf{r})+n\int d\mathbf{r}'c(|\mathbf{r}-\mathbf{r}'|)h(\mathbf{r}'), \tag{12}$$

with n the number density. This equation, however, contains two unknown quantities $h(\mathbf{r})$ and $c(\mathbf{r})$. To solve it, one need a closure relation between these two quantities. In this work, we choose the well-known HMSA method [31], and write [35]

$$g^{HMSA}(\mathbf{r}) = \exp\{-\beta U_0(\mathbf{r})\} \times \left\{1 + \frac{\exp\{f(\mathbf{r})[\gamma(\mathbf{r}) - U(\mathbf{r})]\} - 1}{f(\mathbf{r})}\right\}, \quad (13)$$

where the interaction potential is divided into short-range part, $U_0(\mathbf{r})$, and long-range one, $U(\mathbf{r})$. The former is a hard-sphere potential, and the second one is the repulsive potential derived in the last section, relation (11). There, the function $\gamma(\mathbf{r})$ represents the difference between the total and direct correlation functions, that is $\gamma(\mathbf{r}) = h(\mathbf{r}) - c(\mathbf{r})$. In equality (13), the quantity $f(\mathbf{r})$ represents the mixing function [31]. For convenience, we adopt for this function that form proposed by Bretonnet and Jakse [36]. The virtue of such a form is that, it ensures the thermodynamic consistency in calculating the internal compressibility by two different ways. The form proposed by the authors is [36]

$$f(\mathbf{r}) = f_0 + (1 - f_0)e^{-1/r}. \quad (14)$$

Here, $f_0$ accounts for the interpolation constant, which is an adjustable parameter, such that $0 \leq f_0 \leq 1$. This constant that serves to eliminate the incoherence thermodynamic, can be fixed equating the isotherm compressibility deduced from virial pressure to that calculated from the zero-scattering angle limit of the structure factor, that is $S(0) = nk_B T \kappa_T$.

In Fig. 3, we report the computed structure factor $S(q)$, which is the Fourier transform of the total correlation function $h(\mathbf{r}) = g(\mathbf{r}) - 1$, versus the dimensionless wavenumber $q\sigma_0$, for 4 values of the reduced number density $n/n^*$. Here, $\sigma_0$ denotes the hard-sphere diameter. These curves clearly show that, the height of the principal peak decreases with increasing number density. As a matter of fact, within more concentrated solutions, the effective interactions between the clothed particles are considerably reduced. In other word, the system tends to its ideal nature. Also, the position of this peak is shifted towards its higher values (small wavelengths), as the number density increases. Finally, remark that the value of the structure factor at zero-scattering-angle $S(0)$ becomes more important with increasing number density.

Having described the structure of the considered solution, it remains us to extract the corresponding thermodynamical properties, within the framework of the HMSA. We are interested in three physical quantities, which are the isotherm compressibility, $\kappa_T$, pressure P, and energy E. The first quantity can be directly obtained from the value of the structure factor at vanishing wave-vector, that is $\kappa_T = S(0)/nk_BT$. The two last can be extracted from the standard integral relations, which involve the pair-correlation function and the interaction potential [28]. These quantities are shown in Table I, for 4 values of the reduced number density $n/n^*$. Here, we have considered the dimensionless quantities $\kappa_T/\kappa_T^0$, $P/P_0$ and the difference $\Delta E = (E - E_0)/Mk_BT$. Here, $\kappa_T^0 = 1/Mk_BT$, $P_0 = nk_BT$, and $E_0 = 3Mk_BT/2$ are their homologous for an ideal gas of M particles. The table I indicates that, as the number density is augmented, the isotherm compressibility increases, but the pressure and energy decrease. Probably, this is due to the fact that the diffusion of the clothed particles is *less* important for high densities, where each core is trapped inside a volume of the order of $\sigma^3$, where $\sigma$ is the size of the inner region around the core, relation (8).

## IV. CONCLUSIONS

We recall that the present paper was concerned with semi-dilute solutions of identical star-polymers (clothed particles). The first purpose is the computation of the effective pair-potential experienced by the star-polymers, which originates from the existence of the excluded volume forces between monomers. We found that the pair-potential is essentially *logarithmic*, below some known characteristic distance, and decays *exponentially* at high distances. Contrary to dilute solutions, the potential explicitly depends on the monomer concentration. Second, with the help of this effective potential and the use of the HMSA integral equation method, we studied the structure and thermodynamics of the considered solution.

We emphasize that the HMSA method used here, is more reliable than the *variational* one. To apply the latter, one needs a reference system. For fluids and colloids, for instance, one often chooses the hard-sphere system as reference, where the colloids are considered as impenetrable spheres repelling each other through a hard-core potential. For this case, the variational parameter may be the hard-sphere diameter. In spite of its simplicity, the weakness of the variational method is its mean-field character.

We point out that the choice of the characteristic particle density $n^* = 1/R_G^3$ solves the ambiguities concerning the structure factor, for higher values of the reduced density $n/n^*$.

Finally, the results presented in this paper can be regarded as a natural extension of those relative to dilute solutions. A possible extension to ternary polymer solutions, made of two chemically incompatible polymers in a good solvent, would be an interesting problem to undertake.

**ACKNOWLEDGMENTS**

We are much indebted to Professors J.-L. Bretonnet, J.-M. Bomont and N. Jakse for helpful discussions. Three of us (M.B., F.B. and A.D.) would like to thank the *Laboratoire de Théorie de la Matière Condensée* (*Metz University*) for their kinds of hospitality during their regular visits.

# TABLE CAPTIONS

**TAB I :** Typical values of isotherm compressibility, pressure and energy, for 4 reduced particle densities.

## FIGURE CAPTIONS

**FIG. 1 :** Ratio $\sigma/R_G$ versus the dimensionless variable $\phi_0/\phi_0^*$.

**FIG. 2 :** Reduced pair-potential $U(r)/k_B T$ versus the renormalized distance $r/R_G$, for 4 values of the reduced number density.

**FIG. 3 :** Structure factor $S(q)$ from HMSA versus the dimensionless wavenumber $q\sigma_0$, for 4 values of the reduced number density.

| $n/n^*$ | $\kappa_T / \kappa_T^0$ | $P/P_0$ | $(E-E_0)/Mk_BT$ |
|---|---|---|---|
| 1.2 | 0.018 | 20.077 | 12.912 |
| 2 | 0.037 | 9.277 | 5.130 |
| 2.5 | 0.037 | 9.668 | 5.41 |
| 4 | 0.09 | 4.12 | 1.12 |

**Table I**

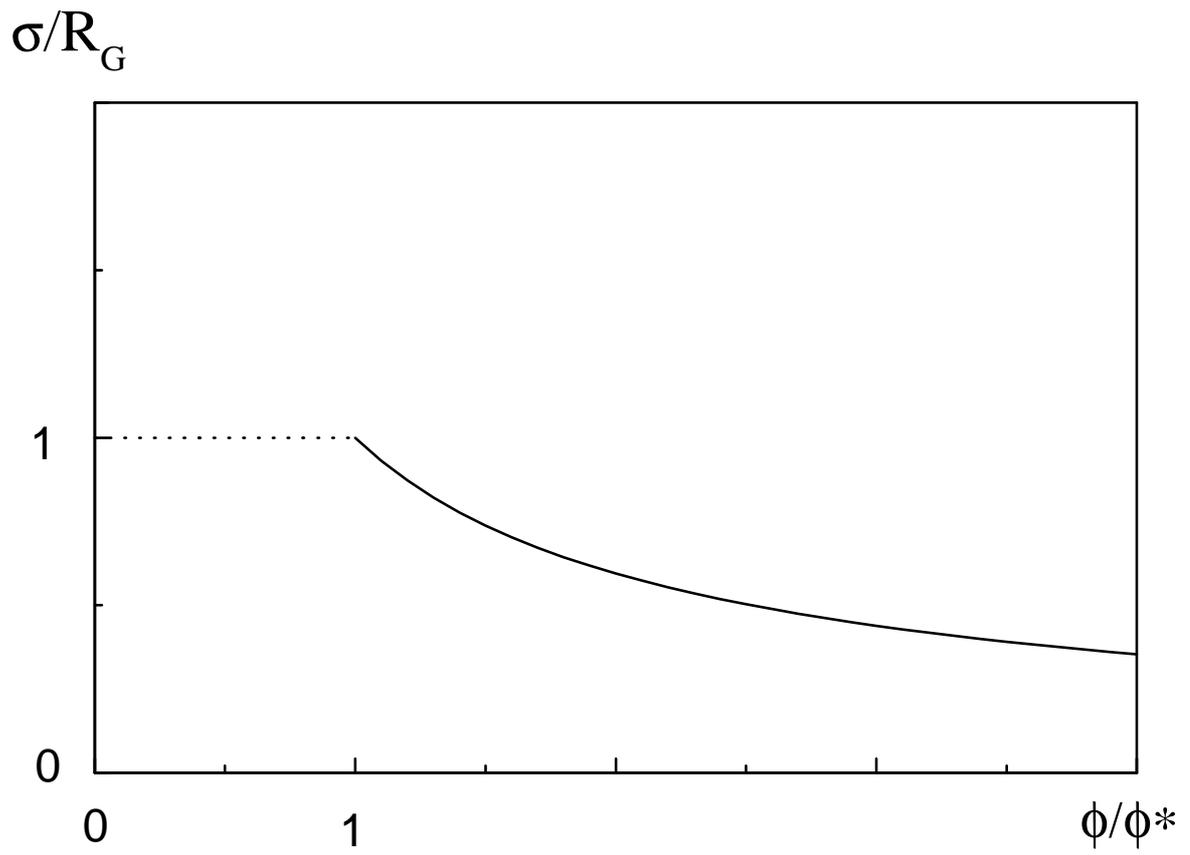

**Figure 1**

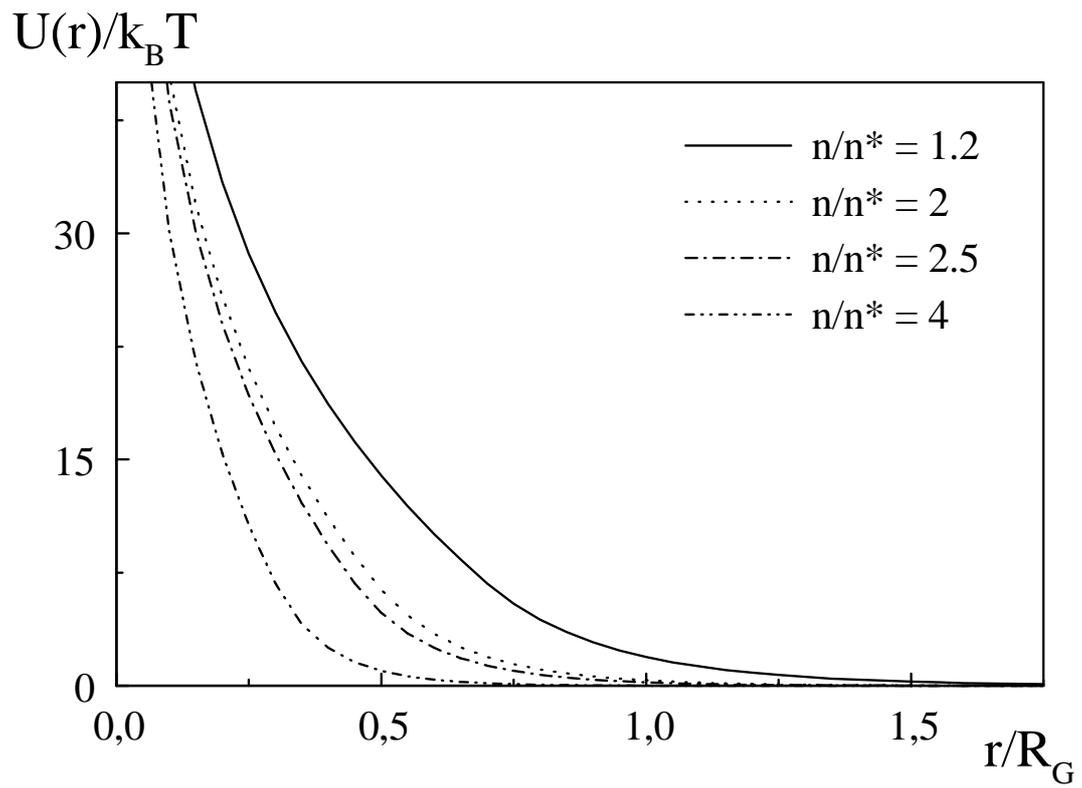

**Figure 2**

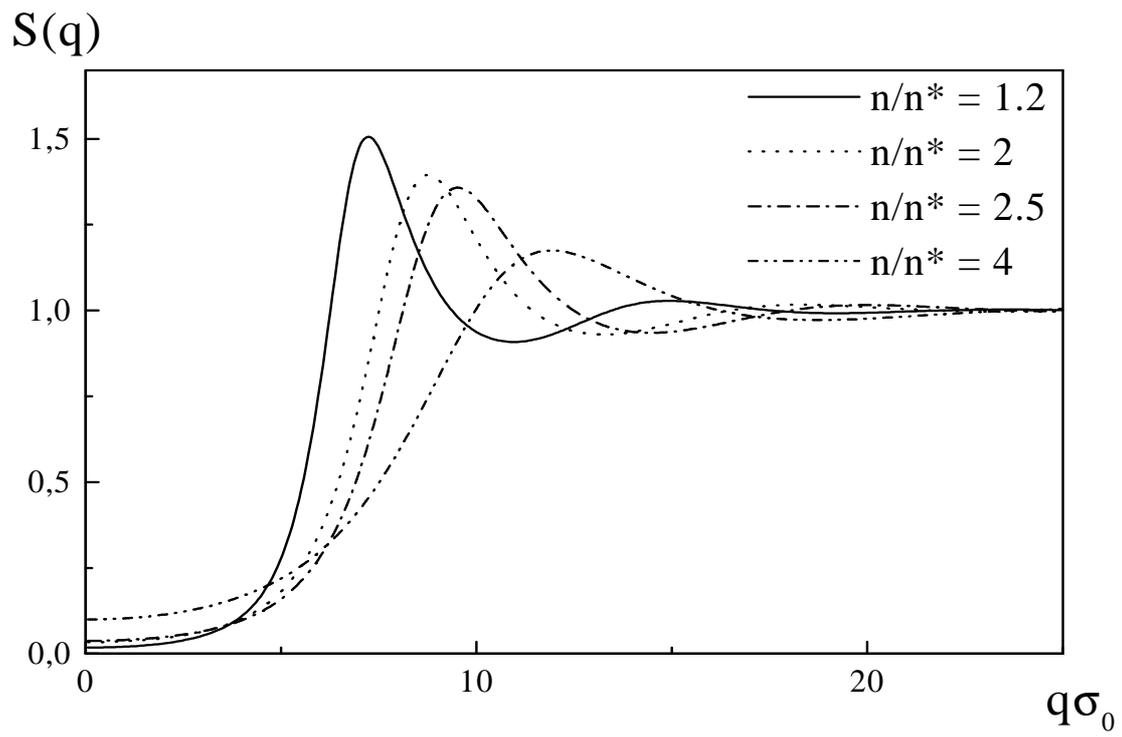

**Figure 3**